\documentclass[fleqn,10pt]{wlscirep}
\usepackage[utf8]{inputenc}
\usepackage[T1]{fontenc}
\usepackage{lineno}
\usepackage{forest}
\usepackage{dirtree}
\usepackage{comment}
\definecolor{folderbg}{RGB}{124,166,198}
\definecolor{folderborder}{RGB}{110,144,169}
\def\Size{4pt}
\tikzset{
  folder/.pic={
    \filldraw[draw=folderborder,top color=folderbg!50,bottom color=folderbg]
      (-1.05*\Size,0.2\Size+5pt) rectangle ++(.75*\Size,-0.2\Size-5pt);  
    \filldraw[draw=folderborder,top color=folderbg!50,bottom color=folderbg]
      (-1.15*\Size,-\Size) rectangle (1.15*\Size,\Size);
  }
}

\title{Tropical Cyclone-Following Dataset for Limited-Area Machine Learning Applications}
\author[1$\dag$]{Minh-Khanh Luong}
\author[1,*$\dag$]{Chanh Kieu}
\affil[1]{Department of Earth and Atmospheric Sciences, Indiana University, Bloomington, IN 47405.}
\affil[*]{Corresponding author: M.K. Luong (mk223338@gmail.com), C. Kieu (ckieu@iu.edu)}
\affil[$\dag$]{Both authors contributed equally to this work.}
\begin{abstract}
This work presents a lightweight dataset of tropical cyclone (TC)-following meteorological variables to facilitate the development and application of machine learning (ML) methods in TC research. Using outputs from our previous climate downscaling and idealized simulations, the dataset is constructed as multichannel video sequences that track TC centers throughout their lifetimes. To demonstrate the dataset’s utility, we present an application in which a regional ML model is trained to predict TC intensity with explicit incorporation of lateral boundary conditions. Our results highlight the importance of lateral boundary information for predicting TC intensity with ML models inside a limited-area domain, underscoring the need to properly account for lateral boundaries when developing regional ML models for TC intensity forecast. 
\end{abstract}

\begin{document}
\flushbottom
\maketitle
\thispagestyle{empty}
%
%
\section{Background} 
Tropical cyclone (TC) intensity dictates the severity of damaging winds, storm surges, extreme precipitation, coastal flooding, and associated socioeconomic losses, yet accurate prediction of TC intensity remains a challenging task in operation. While substantial progress has been achieved in forecasting TC tracks over the past several decades, improvements in intensity prediction have been considerably slower, making it a high-priority area for both research and operational forecasting (HFIP\cite{Gall_etal2013hfip, Franklin_etal2003, Aberson_etal2023, Kieu_etal2021coamps, Poyer_etal2023, Poyer_etal2023}).

Unlike TC motion that is mostly controlled by large-scale steering flow, TC intensity changes arise not only from its internal dynamics but also from their complex multiscale interactions with ambient environment, ocean, and land surface. In particular, TC dynamics are governed by various nonlinear processes operating across a broad range of spatial and temporal scales \cite{Black_etal2002, Gopalakrishnan_etal2011, Judt_etal2016, KieuRotunno2022}. These processes are sensitive to environmental factors such as vertical wind shear, sea surface temperature, or environmental stability, which can be rapidly amplified through nonlinear feedback mechanisms. Thus, current TC forecasting models still exhibit substantial errors, especially during episodes of rapid intensification or rapid weakening \cite{DeMaria_etal2007, Tallapragada_etal2015, Davis_etal2011, DeMaria2010, KieuMoon2016, Kieu_etal2021coamps}.


Over the past few years, explosive machine learning (ML) methods have emerged for TC research \cite{liang_etal2024, miller_etal2017, Olander_etal2021, Wimmers_etal2019}. Deep learning architectures such as convolutional neural networks, recurrent neural networks, transformers, or graph neural networks have demonstrated the ability to extract complex nonlinear relationships from large observational and reanalysis datasets. Many studies have successfully incorporated satellite imagery, infrared brightness temperatures, microwave observations, environmental reanalysis fields, and historical best-track data to improve short-term estimates of TC intensity or identify rapid intensification events. These data-driven methods appear to outperform traditional statistical models for specific tasks when sufficient training data are available.

Specific to TC intensity forecast, current global AI weather models can provide medium-range predictions of TC track and intensity several days in advance and demonstrating skill comparable to, or in some cases exceeding, leading physics-based forecasting systems. For example, the experimental real-time TC forecast by Google DeepMind produces ensemble predictions of TC track and intensity up to five days in advance and has demonstrated competitive prediction relative to operational physics-based models \cite{Ferran2026}. Nevertheless, important challenges remain such as representing the detailed inner-core processes, maintaining physically consistent structures, or quantifying forecast uncertainty, which are more apparent for extreme cases such as Hurricane Polo (2026). 

Another limitation is that existing AI models rely on heterogeneous datasets assembled from multiple observational platforms, satellite sensors, or operational products with inconsistent spatial resolutions, temporal frequencies, and preprocessing procedures. In fact, most available datasets are designed primarily for TC tracking, classification, or satellite image interpretation rather than for representing the complete temporal evolution of TC three-dimensional inner-core details. Such inconsistencies complicate model comparisons, hinder reproducibility, and constrain the development of generalized data-driven forecasting frameworks. 

These limitations highlight the need for a dedicated dataset specifically designed to investigate TC evolution. Such a dataset should provide temporally continuous, spatially consistent, and physically relevant representations of the environmental conditions governing TC development throughout TC lifecycle. In this study, we will present a TC-following dataset to fill this gap. 
Rather than aiming to provide a unique or comprehensive dataset, our main focus is to provide a lightweight dataset that allows researchers to develop ML applications for TC research without the need for handling various nonhomogeneous observational data or extensive physics-based simulations to generate training data. Our multichannel, video-like dataset thus offers a quick benchmark for comparing ML algorithms, improve reproducibility, and serve as a reference for developing AI-based TC intensity forecasting within a regional modeling framework.
%
%
\section{Dataset description}\label{sec:description}
\subsection{Data structure overview}
To ensure reproducibility and facilitate current ML workflows/libraries, our TC-following dataset is organized into a structure that separates data input sources and processing stages. Figure \ref{fig:dataset_tree} shows the dataset directory structure that consists of both the dataset and related scripts to build the dataset from either idealized simulation or real-environment simulations. Specifically in this study, we provide by default the TC-following datasets obtained from the simulation output of the Weather Research and Forecasting (WRF) model, using the idealized high-resolution simulations and the Coupled Model Intercomparison Project Phase 6 (CMIP6) historical downscaling from 1979-2015 and future downscaling (SSP585) from 2080-2100 for the Northwestern Pacific basin \cite{Vu_etal2024, Kieu_etal2025a}.   


\begin{figure}[htbp]
\centering
\small
\begin{minipage}{0.9\textwidth}
\dirtree{%
.1 {./tc-following-dataset}.
.2 {docs/}.
.2 {\fcolorbox{black}{lightgray}{dataset}/}.
.3 {cmip6/}.
.4 {level\_1\_data}.
.4 {\fcolorbox{black}{lightgray}{level\_2\_data}}.
.5 \fcolorbox{black}{lightgray}{cmip\_\{num\_frame\}f\_\{num\_levels\}v\_X.npy}.
.5 \fcolorbox{black}{lightgray}{cmip\_\{num\_frame\}f\_\{num\_levels\}v\_Z.npy}.
.3 {idealized/}.
.4 {level\_1\_data}.
.4 {\fcolorbox{black}{lightgray}{level\_2\_data}}.
.5 \fcolorbox{black}{lightgray}{ideal\_\{num\_frame\}f\_\{num\_levels\}v\_X.npy}.
.5 \fcolorbox{black}{lightgray}{ideal\_\{num\_frame\}f\_\{num\_levels\}v\_Z.npy}.
.2 {input/}.
.3 {best\_track/}.
.3 {cmip6/}.
.3 {wrf/}.
.2 {output/}.
.2 {src/}.
.3 {config.yaml}.
.3 {config\_utils.py}.
.3 {extractor.py}.
.3 {job.sh}.
.3 {step1\_cropping.py}.
.3 {step2\_merging.py}.
.2 {requirements.txt}.
.2 {README.md}.
}
\end{minipage}
\caption{Directory structure and description of the TC-following dataset workflow. Shaded boxes highlight the components of the dataset generated specifically for this study and publicly available on Huggingface repository (see Section 4 for the full data link). The remaining components represent the scripts required to reproduce or generate a similar TC-following dataset. Detailed instructions for using these codes are provided in Section 2.2, and the data access is provided in Section 4.}
\label{fig:dataset_tree}
\end{figure}

For the dataset derived from CMIP6 downscaling, all TCs identified throughout the entire downscaling period are organized into a single spatiotemporal tensor with dimensions $[N, F,H,W,C]$, where $N$ is the number of TCs, $F$ is the number of frames (i.e., the number of output during the lifetime of a TC), $(H,W)$ are the domain size in the latitudinal and longitudinal directions, and $C$ is the number of channels.  This data structure is analogous to the conventional video representation commonly used in image and signal processing. The entire dataset is stored in a NumPy array file named \texttt{cmip\_\{F\}f\_\{C\}v\_X.npy} to reduce data-loading overhead. The corresponding TC metadata are stored separately in another NumPy file \texttt{cmip\_\{F\}f\_\{C\}v\_Z.npy} as shown in Fig. \ref{fig:dataset_tree}.     

Similar to the CMIP6-derived data, the dataset for idealized simulations can be carried out with a naming convention of \texttt{ideal\_\{F\}f\_\{C\}v\_X.npy}, with corresponding auxiliary arrays \texttt{Z} containing all metadata. Unlike the CMIP6-derived data, the idealized dataset includes 12 key variables most relevant for TC intensity and structure retrieval instead of 14 variables. This subset of variables was selected based on our previous experiments to optimize the retrieval of TC intensity and structure \cite{Kieu_etal2025b}. The use of such a reduced number of channels also helps minimize the storage footprint, improve data-access efficiency, and facilitate ML development and large-scale testing. The resulting datasets are stored as NumPy arrays that can be loaded directly using \texttt{NumPy}, enabling efficient downstream model training and evaluation. 

Due to hosting limitations imposed by Hugging Face, only the final (Level-2) dataset shown in Fig. \ref{fig:dataset_tree} are included in this release. All raw CMIP6 and idealized simulation outputs as well as the intermediate (Level-1) data are not provided.

\subsection{Data generation}
To help users to reproduce our dataset or construct new datasets from different input sources, we provide in this section a detailed description of the procedures used in our workflow. Basically, generating the TC-following dataset requires two main components: 1) input data from some physics-based model simulations, and 2) a procedure for constructing the TC-following dataset. Details of each component are provided below:

\subsubsection{Input data preparation}
In our design, we support two main types of input data, both consisting of gridded output from WRF simulations. The first consists of real-environment simulations such as climate downscaling or real-time simulations in which TCs move with the environmental steering flow. The second consists of idealized simulations in which TCs remain near the center of the model domain throughout the simulation. These two categories encompass most common types of TC simulations for practical applications, and we therefore limit the current dataset framework to these two input types.

\begin{itemize}
\item \textit{Real environments}: 
The main characteristic of this input data type is that TCs move with the environmental steering flow, such that their intensity evolves in response not only to internal storm dynamics but also to changes in the surrounding environment. For this input type, we use our previous CMIP6 downscaling products generated by the WRF model, as described in \cite{Vu_etal2024, Kieu_etal2025a}. These downscaling simulations were conducted using the bias-corrected CMIP6 data of \cite{Xu_etal2021cmip6} for both historical and future climate periods. The dataset includes three scenarios: a historical period spanning (1979-2005), and late-century projections under SSP245 and SSP585 for the period 2080-2100. Together, these scenarios provide a baseline representation of the present-day climate and two contrasting future climate states.

As described in Vu et al. (2024, \cite{Vu_etal2024}), the WRF model was configured with a single domain at a horizontal grid spacing of 9 km, consisting of $720\times900$ grid points in the meridional and zonal directions, respectively that covers most of the Western North Pacific (WNP). Model outputs were archived at 6-hour intervals. Because approximately 25-30 TCs occur annually in the WNP, TC-active periods represent only a small fraction of the total simulation period. Nevertheless, with 20 years of historical simulations and 20 years of future simulations for the late-century period, our CMIP6 downscaling dataset provides a sufficiently large number of TC cases for constructing the TC-following dataset included in this release.

\item \textit{Idealized environments:}
In addition to the real-environment simulations that track individual TCs, we conducted a range of idealized WRF simulations designed to isolate the sensitivity of TC evolution to different physical parameterizations. Specifically, a set of numerical experiments were initialized from either different initial conditions or physics parameterization schemes. The model domain was fixed (i.e., non-moving) on an $f$-plane centered at $20^\circ N$ such that the simulated TC remained at the center of the domain throughout the integrations. These WRF idealized experiments were configured with a nested domain of horizontal resolution 18/6/2 km, and model outputs were archived at 10-minute intervals. The vertical grid consists of 41 levels spanning from 1000 to 30 hPa. Additional configuration details, including the model time step, model-top pressure, and specific physics options, are provided in \cite{Kieu_etal2025b}.
\end{itemize}

\subsubsection{Construction procedures}
Given the two different types of input data used to generate the TC-following dataset as discussed above, different preprocessing approaches are required for the different input data types. 

For the idealized simulations in which the TC center remains fixed at the domain center throughout the simulation, WRF outputs were stored at their native temporal resolution in individual files, with each file containing a single time step. For each registered TC event, a $100 \times 100$ pixel domain was extracted at the corresponding time step, retaining the complete, unprocessed model output within the selected region.

For the CMIP6 downscaled data, we employ a physics-based TC tracking algorithm following Vu et al. (2021, \cite{Vu_etal2021}), which applies dynamically and thermodynamically consistent criteria to the WRF output fields. To maintain a TC-centered reference frame, the native WRF coordinate system is transformed into a storm-relative moving coordinate framework. In addition, the original latitude and longitude coordinates are retained as time-dependent two-dimensional fields for geophysical reference.

With such different requirements for preprocessing steps, our dataset-generation workflow organizes the procedures according to input data source and processing level, with separate \texttt{cmip} and \texttt{idealize} configurations (Fig. \ref{fig:dataset_tree}). Within each input data type, scripts with prefix \texttt{step1} generate storm-specific data NetCDF files (L1), while scripts with prefix \texttt{step2} merge all storms into a single ML-ready NumPy array (L2). Details of these procedures are provided below:

\paragraph{Step 1 - Cropping TC domain:}
For the CMIP6 data, \texttt{step1\_cropping\_cmip6.py} groups all TC records according to storm identifiers output from a vortex tracking algorithm. Since our dataset follows TC centers at 6-hour interval, the local spatial dimensions are renamed \texttt{x} and \texttt{y}, while the original geographic coordinates are retained as \texttt{TRUE\_LONG} and \texttt{TRUE\_LAT}. In addition, the storm-center coordinates are stored as the time-dependent variables \texttt{cen\_lon} and \texttt{cen\_lat} in the metadata. This CMIP6 workflow also records \texttt{storm\_id} and the storm lifetime \texttt{lifetime\_frame}. All time steps associated with a given storm are then concatenated into a single NetCDF file for each TC.

For the idealized data, \texttt{step1\_cropping\_idealized.py} handles a bit differently, as it processes the entire experiment directories listed in \texttt{config.yaml}. This script sorts input files whose names contain the domain to be processed (\texttt{d01} by default), extracts a centered $100\times100$ grid-point window, and sorts the concatenated records chronologically. The domain center is fixed in all idealized experiments and so it does not contain any geographic information.

Note that the WRF model employs a staggered grid, in which some variables are defined at locations offset from the primary mass grid. To obtain a spatially consistent representation, all staggered fields are linearly interpolated onto a unified mass grid (A-grid) before subsequent step. 


\begin{table}[ht]
\centering
\caption{Meteorological variables included in the TC-following dataset.}
\label{tab:input_vars}
\begin{tabular}{|c|l|l|l|}
\hline
\textbf{Channel} & \textbf{CMIP6} & \textbf{Idealized} & \textbf{Description} \\
\hline
1  & \texttt{U10m}      & \texttt{U10m} & U wind at 10 m (m s$^{-1}$) \\
2  & \texttt{V10m}      & \texttt{V10m} & V wind at 10 m (m s$^{-1}$) \\
3  & \texttt{SST}       & \texttt{SST}  & Sea surface temperature (K) \\
4  & \texttt{LANDMASK}  & \texttt{LANDMASK} & Land sea mask (0/1)\\
5  & \texttt{U28}       & \texttt{U14} & 100 hPa upper-level U wind (ms$^{-1}$)\\
6  & \texttt{V28}       & \texttt{V14} & 100 hPa upper-level V wind (ms$^{-1}$) \\
7  & \texttt{U05}       & \texttt{U03} & Boundary layer U wind (ms$^{-1}$) \\
8  & \texttt{V05}       & \texttt{V03} & Boundary layer V wind (ms$^{-1}$) \\
9  & \texttt{T23}       & \texttt{T12} & 300-hPa temperature (K) \\
10 & \texttt{QVAPOR10}  & \texttt{QVAPOR05} & 850-hPa water vapor (g/kg) \\
11 & \texttt{PHB10}     & \texttt{PHB05} & Geopotential height at 800 hPa (m)\\
12 & \texttt{PSFC}      & \texttt{SLVL} & Sea level pressure (hPa) \\
13 & \texttt{RAINNC}    & -- & Non-convective rainfall rate (mm/hr) \\
14 & \texttt{RAINC}     & -- & Convective rainfall rate (mm/hr) \\
\hline
\end{tabular}
\end{table}

\paragraph{Step 2 - Merging TC domain:}
With the L1 outputs generated, the scripts \texttt{step2\_merging\_cmip6.py} and \\ \texttt{step2\_merging\_idealized.py} finally construct fixed-length samples according to the corresponding data type. Let $N$ denote the total number of samples, $F$ is the number of \texttt{frames}, $S$ the value of \texttt{sample\_stride}, $H$ and $W$ the spatial dimensions of the cropped domain, and $C$ the number of channels. The resulting L2 output array will then have a shape of $(N,F,H,W,C)$.

For the released dataset, the current configurations for both the CMIP6 and idealized WRF simulations use $F=9$, $S=4$, $H=W=100$, and $C=12$ for idealized simulations or 14 for CMIP6 downscaling (see the list of input channels in Table \ref{tab:input_vars}). These variables are selected to represent the multilevel dynamical and thermodynamic structures of the TCs and their surrounding environments. Specifically, the selected fields characterize near-surface flow, boundary-layer dynamics, mid-tropospheric thermodynamic conditions, upper-level outflow, moisture distribution, and surface characteristics, all of which are important for TC structure and intensity evolution. 

Along with the main data, an auxiliary metadata array $Z$ has the shape of $(N,F,6)$. Its dimension is ordered as \texttt{[cen\_lon, cen\_lat, sin, cos, storm\_lifetime, start\_frame]}. Here, \texttt{sin} and \texttt{cos} denote the cyclic day-of-year features $\sin(2\pi,\mathrm{DOY}/Y)$ and $\cos(2\pi,\mathrm{DOY}/Y)$, respectively, where $Y=365$ or $366$. Here, \texttt{start\_frame} denotes the starting index for each sequence, which is repeated across the $F$ frames of each sample.
Figure \ref{fig:dataset_tree} summarizes all details of the two-level dataset procedures.

%
%
\subsection{Dataset validation}
\begin{figure}[t]
\centering
\includegraphics[width=1\linewidth]{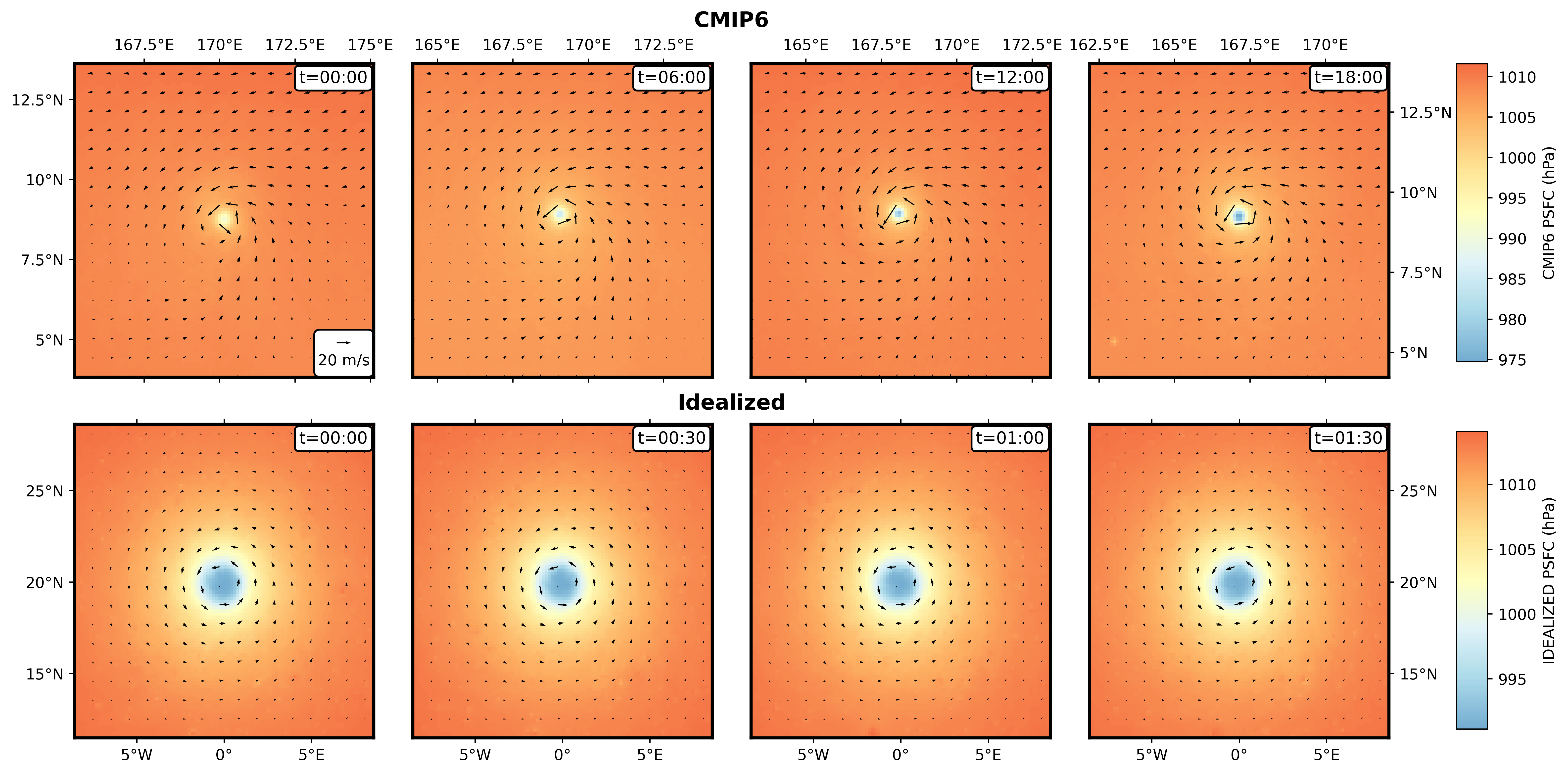}
\caption{An illustration of four consecutive time frames of TC fields including sea level pressure (shaded, unit mb) and 10-m wind (vectors) obtained from the CMIP6 dataset (top panels) and idealized TC-following dataset (bottom panels)}
\label{fig:val_cmip_idealize_sequence}
\end{figure}

To validate our dataset, the most direct approach is to examine a sequence of data frames for individual storms. For this purpose, an illustrative sequences from the CMIP6 and idealized WRF datasets are shown in Fig. \ref{fig:val_cmip_idealize_sequence} over consecutive time steps. One can see that the consecutive frames remain physically consistent throughout the dataset construction process, with the temporal evolution of the fields preserved without altering the underlying dynamical structures. For example, the sea level pressure field clearly exhibits the intended TC-relative moving coordinate framework. In this regard, Fig. \ref{fig:val_cmip_idealize_sequence} provides a direct verification of the structural and physical consistency of the constructed dataset. 

The storm-centered pressure minimum and coherent cyclonic circulation also remain well preserved, confirming that the coordinate transformation and staggered-to-mass-grid interpolation introduce no apparent spatial distortion. The smooth evolution across consecutive frames further demonstrates the temporal integrity of the preprocessing pipeline. The consistent structures observed in both the CMIP6 and idealized datasets indicate that the dataset construction procedure preserves the underlying physical fields without introducing noticeable processing artifacts.
%
%
\section{Application: TC intensity prediction}
To illustrate how the above TC-following dataset can be used in TC research, we present in this section an ML development that predicts the next TC state within a limited-area domain, based on past TC states and lateral boundary conditions (LBC). As in physics-based models, future lateral boundaries play a critical role in determining the evolution of systems within the domain in ML models designed for limited-area applications \cite{Oskarsson_etal2023graph, Abdi_etal2026hrrrcast, Adamov_etal2025graph, Larsson_etal2025diffusionlam}. This is important for TC intensity prediction, as the lateral boundaries provide updated information about the surrounding environment, which can modulate TC intensity as strongly as the internal vortex dynamics.     

Specific to our ML model here, let $\mathbf{X}_t \in \mathbb{R}^{H \times W \times C}$ denote the TC state at time $t$, $\Delta t$ the time step, $\tau$ the number of input frames, and $\mathbf{Z}_(t+\Delta t) \in \mathbb{R}^{H \times W \times C}$ the LBC at time $t + \Delta t$. Our DL operator to predict the next TC state at time $t+\Delta t$ is defined as
\begin{equation}
\hat{\mathbf{X}}_{t+\Delta t} = h\left(\mathbf{X}_{t-\tau\Delta t}, \ldots, \mathbf{X}_t, \mathbf{Z}_{t+\Delta t}\right),
\end{equation}
where the objective is to learn $h$ such that $\hat{\mathbf{X}}_{t+\Delta t} \approx \mathbf{X}_{t+\Delta t}$. This operator differs from statistical downscaling in that it learns a temporal evolution operator conditioned on past TC states and future LBC, rather than performing an instantaneous mapping from coarse to high-resolution grid. Compared to dynamical downscaling, which explicitly solves governing physical equations, our approach learns an implicit, data-driven approximation of TC evolution, thus offering a computationally efficient surrogate while retaining key dynamical dependencies.

\begin{figure*}[t]
\centering
\includegraphics[width=\textwidth]{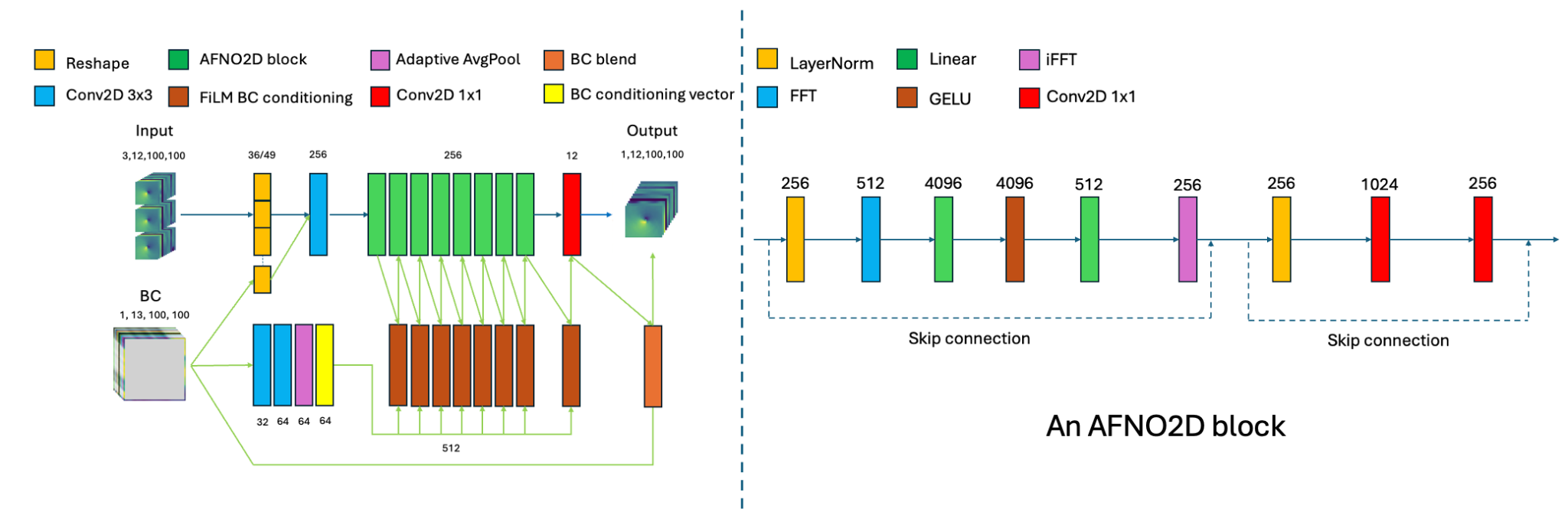}
\caption{Overview of the proposed model architecture based on the Adaptive Fourier Neural Operator (AFNO) architecture. The network takes TC-centered past fields up to time $t$ and lateral boundary conditions at time $t+1$ as inputs to predic the full field at time $t+1$. The main branch processes TC fields through convolutional layers followed by multiple AFNO2D blocks to capture spatiotemporal structures, while a parallel conditioning branch encodes LBC information via convolution and adaptive pooling. The LBC features are injected into the AFNO blocks using three different mechanisms shown by green arrows, which include 1) FiLM-based conditioning, 2) direct embedding, and 3) blending into the output through a separate LBC blending module. The AFNO2D block (right) consists of LayerNorm, Fourier transform (FFT/iFFT), channel mixing via linear layers and GELU activation, and residual skip connections.}
\label{fig:model_arch}
\end{figure*}

For illustration purposes, we employ the Adaptive Fourier Neural Operator (AFNO, \cite{Pathak_etal2022, Luong2026MS}) for our DL operator $h$, which is a neural operator that learns mappings between spatial fields in the spectral domain. By operating in the Fourier space, AFNO can efficiently capture long-range dependencies and multi-scale structures that are difficult for purely convolutional models to represent. This property is especially relevant for TC modeling, where storm intensity and structure are governed by interactions between large-scale environmental forcing, such as steering flow, vertical wind shear, and thermodynamic conditions, and fine-scale inner-core processes. Because these interactions span multiple scales and are not purely local, AFNO provides a suitable framework for TC-related prediction tasks. The overall model architecture is summarized in Fig. \ref{fig:model_arch}, with more details in \cite{Luong2026MS}.

The core of AFNO lies in the AFNO2D block, which operates by transforming spatial features into the spectral domain, applying learned transformations, and then projecting them back to physical space. Specifically, the input feature map is first passed through a 2D Fast Fourier Transform (FFT), converting spatial patterns into frequency components where each mode represents a global structure across the domain. In this spectral space, AFNO applies a learnable, adaptive filtering mechanism that selectively modulates important frequency modes while suppressing less relevant ones, effectively learning how different spatial scales contribute to the target dynamics. After this spectral processing, an inverse FFT (IFFT) is applied to map the features back to the spatial domain. The resulting representation is then combined with residual connections and further processed through pointwise feed-forward layers, refining the original details.

We incorporate LBC through three mechanisms. First, a direct input strategy embeds the LBC into the same spatial-channel format as the past frames, enabling the model to explicitly access future environmental information alongside the storm state. Second, a Feature-wise Linear Modulation (FiLM)-based conditioning approach modulates the intermediate feature maps after each AFNO2D block by applying feature-wise affine transformations conditioned on the LBC. Specifically, given an intermediate feature map $\mathbf{F} \in \mathbb{R}^{H \times W \times C}$, FiLM applies
\begin{equation}
\text{FiLM}(\mathbf{F} \mid \mathbf{Z}) = \gamma(\mathbf{Z}) \odot \mathbf{F} + \beta(\mathbf{Z}),
\end{equation}
where $\gamma(\mathbf{Z}) \in \mathbb{R}^{C}$ and $\beta(\mathbf{Z}) \in \mathbb{R}^{C}$ are learned scaling and shifting parameters generated from the LBC through a conditioning network, and $\odot$ denotes channel-wise multiplication. Third, a direct blending method incorporate the LBC into the predicted output field through a weighted combination. While this approach does not influence the internal feature evolution as in the first two methods, it plays a crucial role in stabilizing rollout training and ensuring consistency with boundary constraints during inference.

It should be emphasized again that how to properly handle LBCs is the key for capturing TC evolution, given that TC intensity is strongly influenced by environmental conditions \cite{Du_etal2013, Kieu_etal2021coamps}. Thus, different treatments of LBCs can significantly alter TC evolution pathway. If LBC information is not incorporated, the model effectively assumes a single deterministic mapping in time for all future frames (i.e., nowcasting), which limits its ability to capture environment-dependent variability and leads to degraded performance for longer lead times when rolling out. 

With this limited-area model architecture, we design our baseline experiment using the TC-following dataset generated from the CMIP6 downscaling. The objective is to predict a TC state at a lead time of 6 hours (corresponding to one frame into the future) conditioned on the previous three frames (12 hours of TC history) and the future LBCs. Note that in the spectral domain, the TC central regions are often characterized by enhanced high-frequency components associated with sharp gradients and fine-scale structures, such as spiral rainbands, eyewall features, and compact wind and pressure extrema. In the physical domain, they correspond to the most extreme values of atmospheric variables such as the strongest winds, lowest pressures, and strongest localized thermodynamic anomalies. To better capture these TC unique characteristics, we thus employ a composite loss function as follows:
\begin{equation}
\mathcal{L}_{\text{total}} =
w_{L1}\mathcal{L}_{L1} +
w_{\text{Extreme}}\mathcal{L}_{\text{Extreme}} +
w_{\text{HighFreq}}\mathcal{L}_{\text{HighFreq}}.
\end{equation}
In the baseline experiments, we set
\(
w_{L1}=1.0,\;
w_{\text{Extreme}}=0.5,\;
w_{\text{HighFreq}}=1.0.
\)
The individual loss terms are defined as
\begin{equation}
\mathcal{L}_{L1} =
\mathrm{mean}
\left(
\left|\hat{\mathbf{X}}_{t+\Delta t} - \mathbf{X}_{t+\Delta t}\right|
\right),
\end{equation}
\begin{equation}
\mathcal{L}_{\text{Extreme}} =
\mathrm{mean}
\left(
\tilde{\mathbf{w}}_{\text{ext}}
\odot
\left|\hat{\mathbf{X}}_{t+\Delta t} - \mathbf{X}_{t+\Delta t}\right|
\right),
\end{equation}
\begin{equation}
\mathcal{L}_{\text{HighFreq}} =
\mathrm{mean}
\left(
\left|
\mathcal{F}
\left(
\hat{\mathbf{X}}_{t+\Delta t} - \mathbf{X}_{t+\Delta t}
\right)
\right|
\odot
\mathbf{M}_{\text{hf}}
\right).
\end{equation}
Here, $\mathcal{L}_{L1}$ measures the mean absolute error over the full domain. The extreme loss $\mathcal{L}_{\text{Extreme}}$ applies spatial weights derived from the normalized magnitude of the target field, thereby emphasizing regions with strong extremes such as the TC core. The high-frequency loss $\mathcal{L}_{\text{HighFreq}}$ penalizes residual errors in the high-wavenumber portion of the spectrum, encouraging the model to better reconstruct fine-scale structures and sharp gradients that are typically smoothed by standard pixel-wise losses.

To reduce temporal dependence between the training, validation, and test sets, we employ a temporally structured data split. Because the dataset is organized into time-ordered chunks, we preserve their chronological order rather than randomly assigning nearby sequences to different sets. Specifically, we select a contiguous 10\% segment as the test set and designate the two adjacent 5\% segments immediately before and after it as the validation set, resulting in a validation-test-validation arrangement of 5\%-10\%-5\%. This design limits the mixing of temporally neighboring samples across the validation and test sets and provides a more temporally coherent evaluation. Note however that the extent to which it reduces temporal correlation will depend on the specific temporal dependence of the data and whether individual storms span multiple splits.

The model is trained for 300 epochs using the Adam optimizer with an initial learning rate of 0.001. A learning rate scheduler is applied such that the learning rate is reduced by a factor of 10 if validation performance does not improve for 7 consecutive epochs, with a minimum threshold of $1\times10^{-6}$. To assess the impact of LBCs, we conduct an additional experiment in which the entire LBC-processing pathway is removed. All other implementation details and configuration settings are discussed in \cite{Luong2026MS}.

\begin{figure*}[t]
\centering
\includegraphics[width=1.0\textwidth]{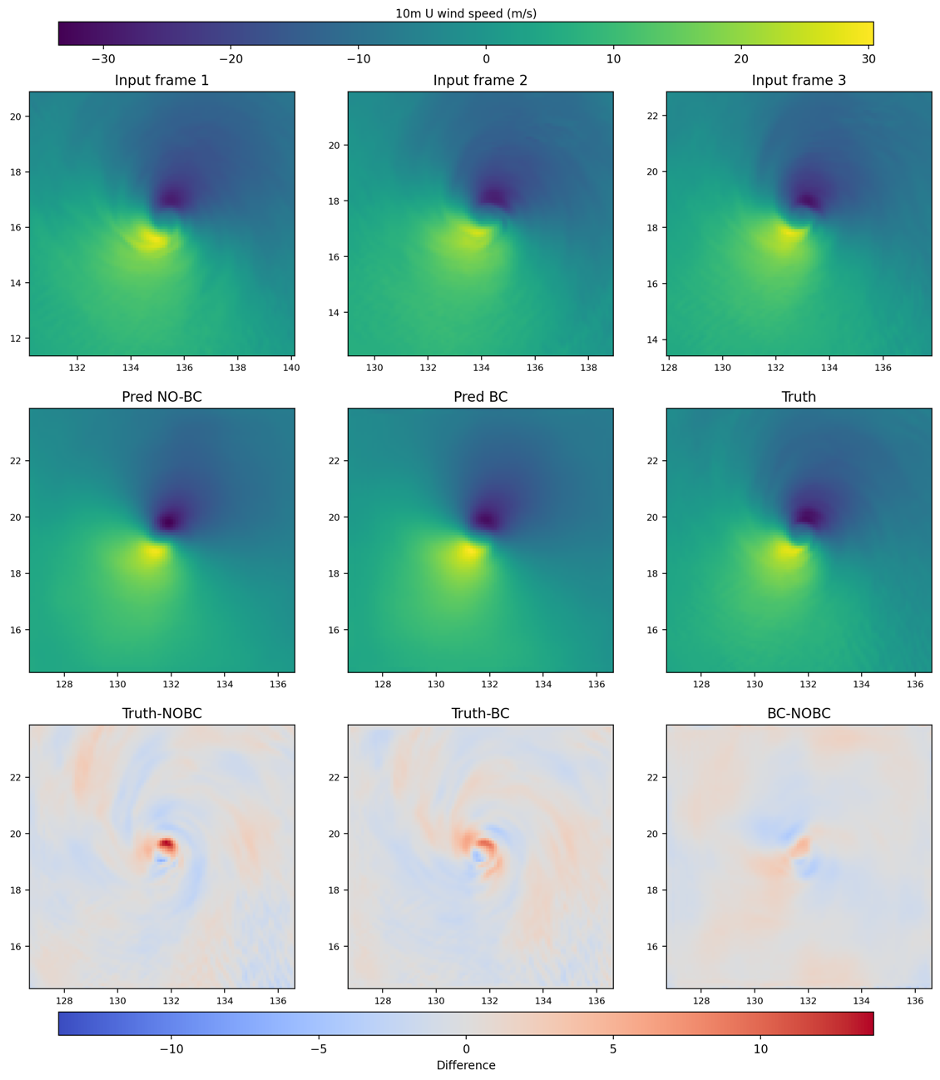}
\caption{Comparison of model predictions for the 10-m U wind field (shaded, unit m s$^{-1}$) with and without LBC conditioning. The top row shows three consecutive input states, covering 12 hours of past history, followed by predictions without LBC (Pred NO-BC, left panel), predictions with LBC (Pred BC, central panel), and the ground truth (right panel). The bottom row shows the difference distribution for Truth-NO BC, Truth-BC, and BC-NO BC.}
\label{fig:BCexample}
\end{figure*}

Fig.~\ref{fig:BCexample} shows an example of the AFNO prediction for one test sample, which captures the overall evolution of TC states as designed. Compared with the prediction without LBCs, the prediction with LBCs better displays the full field distribution of wind, with higher wind-speed peak near the inner core. Such better performance of the AFNO model when including the LBC is seen for other fields and samples, thus supporting that LBCs exert influences on both the TC structure and intensity in our AFNO model. This is a notable feature of the AFNO architecture, as it allows information from the lateral boundaries to influence the TC inner core beyond just the local effects in regions near the boundaries. 

Although not immediately apparent from Fig.~\ref{fig:BCexample}, we note however that some high-frequency components are partially lost in both experiments with and without the LBC. While the model reproduces the overall TC intensity and structural patterns with reasonable fidelity with LBC, such a loss of high-frequency information limits its ability to accurately capture peak intensity for very strong TC cases, indicating some current limitation of our AFNO optimization used in this study.

\begin{figure*}[t]
\centering
\includegraphics[width=\textwidth]{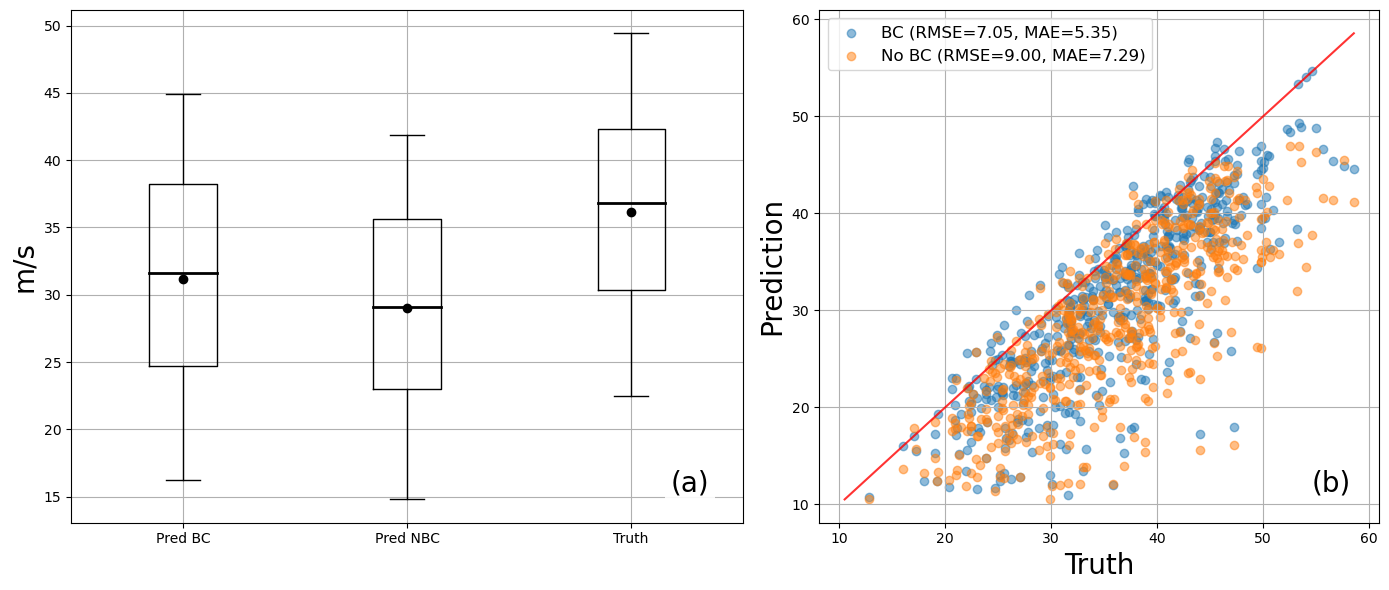}
\caption{(a) Boxplots show the interquartile range, median (line), mean (dot), and whiskers (5-95\%) of the maximum 10-m wind (unit m s${-1}$) for the experiment with LBC (Pred BC) and without LBC (Pred No-BC) as compared to the truth distribution; and (b) the scatter plot of the prediction from the LBC (blue) and no LBC (orange) experiments against the truth.}
\label{fig:minmaxerror}
\end{figure*}

To more systematically see the role of the LBC on the retrieval of TC intensity (measured in terms of the maximum 10-m wind),  Fig.~\ref{fig:minmaxerror}a shows the boxplot distribution of the 10-m wind from both the LBC and no-LBC experiments as obtained from our AFNO model. Here, the 10-m wind speed for both the prediction and the truth is computed as the magnitude of the horizontal wind vector (i.e., $\sqrt{u_{10}^2 + v_{10}^2}$). Although both experiments with and without LBC could not fully capture the entire range of TC intensity as compared to the truth, using LBC significantly improves the model performance across statistical measures such as mean, inter quartiles, or RMSE. This is consistent with the overall spatial distribution of the wind field seen in Fig. \ref{fig:BCexample}, especially for the peak wind near the TC center. 

The scatter plot in Fig.~\ref{fig:minmaxerror}b further highlights the role of LBCs in predicting TC intensity. The LBC experiment outperforms the No-BC experiment, yielding lower RMSE and MAE and better agreement with the strongest observed intensities. Nevertheless, both experiments exhibit a systematic low bias, with RMSEs of approximately 7 ms$^{-1}$ for LBC and 9 ms$^{-1}$ for No-BC. These results indicate that, although incorporating LBC information improves overall predictive performance, the AFNO model still struggles to capture the full range of TC intensities, potentially due to limitations in its architecture or training strategy. Regardless of the details, these results demonstrate that our TC-following dataset can be useful for developing time-dependent DL applications. Beyond TC full-field predictions or intensity retrieval demonstrated above, our dataset can be used for studies that focus on different aspects of TC structure or development such as rapid intensification, intensity changes under different future climates, predictability studies, or regional operational forecasting. 

%
%
\section{Code availability and usage}\label{sec:user_guide}
The complete source code, software environment, configuration files, and example scripts are publicly available. Both the code and dataset presented herein are fully accessible on our GitHub repository at https://github.com/kieucq/tc-following-dataset and Hugging Face repository: https://huggingface.co/datasets/kieucq/tc-following-dataset. All Python codes follow the standard Open Source License. 

\bibliography{reference_khanh, reference_da, reference_dl, reference_mine, reference_misc, reference_prdb, reference_tcClimate, references_tc}

\section*{Acknowledgments} 
This research is partially supported by the NSF grant \# AGS-2309929.

\section*{Author contributions statement}
CK conceived the limited-area ML ideas. KL implemented the workflow and AFNO model development, as well as some model validation. Both authors analyzed the results and wrote the manuscript. 

\section*{Competing interests} 
The authors herein declare no competing interests for this study.

\section*{Figures \& Tables}
\end{document}